\input{aipcheck}
\documentclass[
    ,final            
  ]
  {aipproc}

\layoutstyle{6x9}

\usepackage{color}

\usepackage[utf8x]{inputenc}
\usepackage{graphicx}
\newcommand*{\mpl}{M_{\rm{Pl}}}
\newcommand*{\lag}{\mathcal{L}}
\newcommand*{\si}{{\rm sign}}

\begin{document}
\title{On the dynamics of unified k-essence cosmologies}
\classification{98.80.-k;98.80.Bp;98.80.Cq}
\keywords{Theoretical cosmology, Phase space analysis.}

\author{Josue De-Santiago}{address={Universidad Nacional Aut\'onoma de M\'exico,  04510 M\'exico DF\\ 
Institute of Cosmology $\&$ Gravitation, University of Portsmouth, Dennis Sciama Building, 
Portsmouth, PO1 3FX, United Kingdom.}}

\author{Jorge L. Cervantes-Cota}{address={Depto. de F\'{\i}sica, Instituto Nacional de Investigaciones
Nucleares, A.P. 18-1027,  11801 M\'{e}xico DF.}}


\begin{abstract}
We analyze the phase space of a particular unified model of dark matter, dark energy,  and inflation that 
we recently studied in \cite{DeSantiago:2011qb} whose Lagrangian is of the form $\mathcal{L}(X,\phi)=F(X)-V(\phi)$.  
We show that this model possesses a large set of initial conditions consistent with  a successful  cosmological
model in which an inflationary phase is possible, followed by a matter era to end with dark energy domination. 
In order to understand the success of the model,  we study the general features that unified dark matter (UDM)  
models should comply and then we analyze some particular models and find their constrictions.

\end{abstract}

\maketitle
\section{Introduction}
The current standard model of cosmology is based, on the one hand, on the existence of dark matter as a clustering 
agent to yield both local galactic dynamics and the large scale structure of the Universe and, on the other hand, on 
dark energy as a substance responsible for the 
cosmic accelerated expansion, and they both comprise around $96 \%$ of the matter-energy content  at present.  The 
rest $4 \%$ is mainly in the form of baryons that are important to understand what we actually observe.   
Astonishingly, this cosmological model have received very much support from different cosmological 
probes, such as type Ia supernovae, CMB anisotropies, measurements of Baryon Acoustic Oscillations, galactic 
and cluster dynamics, among others, for a short review see ref. \cite{CervantesCota:2011pn}.   

In spite of the above-mentioned success, the standard model of cosmology relies on the existence of the dark components 
that make possible the desired cosmological dynamics. So far, we do not know, by certain, the origin of dark matter and dark energy,
although well-motived candidates exist for their origin. Given this, one is tempted to look for alternatives to dark matter and/or dark energy in
such a way that the known, correct dynamics of the standard model is recovered and, if possible, models having smoking guns to 
be able to discriminate among them with the observations at hand.    One  of the interesting possibilities that has appeared in recent years is 
that different phenomena such as inflation, dark matter, and dark energy could be due to a single scalar field 
\cite{Liddle:2006qz,Cardenas:2007xh,Panotopoulos:2007ri,Liddle:2008bm,Lin:2009ta,Capozziello:2005tf,Henriques:2009hq}.  A 
practical scheme is to look for toy models that can accomplish the desired dynamics, and in that
sense it is important to understand the key elements of the Lagrangian that play a particular cosmic role.  Different 
works \cite{Liddle:2006qz,Giannakis:2005kr} have analyzed the features that a unified
model should have, and have pointed out the difficulties to build such a single description of the different phenomena.  Recently,  a unified 
k-essence model for a particular Lagrangian
has been proposed \cite{Bose:2008ew}  and later generalized to a whole class of models
in Ref. \cite{DeSantiago:2011qb}. It was shown there that these models work finely to 
achieve a cosmological dynamics that emulates that of the standard model of cosmology including inflation. However, an initial conditions analysis is missing 
and some key features of the success have not yet been investigated.    The aim of the present work is therefore to analyze the 
phase space dynamics of that model to identify  the key elements to later generalize our results to other unified models, such as 
unified dark matter (UDM) models \cite{Scherrer:2004au,Bertacca:2010ct}.

\section{Dynamical analysis of the unification model $\lag = F(X)-V(\phi)$}

We consider here a scalar field $\phi$ with Lagrangian
$\mathcal{L}(X,\phi)=F(X)-V(\phi)$, where the kinetic term $F$ in this generalized class
of models is a function of the canonical kinetic term $X=-g^{\mu\nu}\partial_\mu \phi\partial_\nu \phi/2$. 
The cosmological equations of motion for this field in a flat Friedmann-Robertson-Walker universe are
\begin{equation}\label{fried1}
   H^2=\frac{1}{3\mpl^2} (2XF_X -F + V) 
\end{equation}
and
\begin{equation}\label{cont}
   \frac{d}{dt} (2XF_X-F+V) + 6HXF_X = 0 \,, 
\end{equation}
where $\mpl^2 \equiv 1/8 \pi G$ and equation (\ref{cont}) is the continuity equation.

In Refs. \cite{DeSantiago:2011qb} and \cite{Bose:2008ew} one
particular choice is made for these Lagrangians corresponding to
\begin{eqnarray}
  F(X) &=&  \frac{1}{(2\alpha-1)} \left[ (A X)^{\alpha} 
  - 2\alpha  \alpha_0  \sqrt{AX} \right] + M\,, \label{eq1} \\
  V(\phi)&=&  \frac{1}{2} m^2 \phi^2 \,, \label{eq2}
\end{eqnarray}
where $\alpha=1$ corresponds to de model studied in \cite{Bose:2008ew}.
In those works it is shown that
this scalar field has the interesting properties to emulate the dark matter and dark energy, and
in the very early Universe to drive inflation. To obtain this behaviour several elections of the
constant parameters have to be made, for example for $\alpha=1$ then
\begin{eqnarray}
10^{-48} \mpl^2< &\alpha_0& <10^{-40}\mpl^2 \,, \nonumber \\ \nonumber
m &\sim& 10^{-6} \mpl \,, \\ \nonumber
\alpha_0^2 -M &\sim& 10^{-120} \mpl^4 \,, \\
A &\sim& 1 \,.
  \label{constants}
\end{eqnarray}
A missing part of the study of these models is the analysis of the 
phase space of its solutions. This can tell us whether the solutions 
that have these cosmologically interesting features are feasible to obtain or not. In other words, it tell us
whether the initial conditions of the solutions are generic or have to be fine tuned.
In Ref. \cite{DeSantiago:2012nk} a study on the general features of the phase space for the models
with Lagrangian $\lag =F(X)-V(\phi)$ is presented. Here
however, we will carry out a similar analysis adapted to the particular choice (\ref{eq1},\ref{eq2}).

The continuity equation, assuming $\alpha=1$ becomes
\begin{equation}
   A\ddot{\phi}+ 3HA\dot{\phi} + m^2\phi=3\sqrt{2A} \alpha_0 \si (\dot{\phi}) H \,,
\end{equation}
where $\si (\dot{\phi})$ is $-1$ if $\dot{\phi}$ is negative and $+1$
otherwise. Performing a change of variable to
\begin{equation}
z  \equiv \sqrt{A}\dot{\phi}\,,
\end{equation}
and using the Friedmann equation (\ref{fried1}) to substitute the value of $H$
\begin{equation}
  H=\sqrt{\frac{-2M+z^2+m^2\phi^2}{6\mpl^2}} \,,
\end{equation}
we obtain an evolution equation for $z$ in terms of the variables $\phi$ and
$z$. This equation and the definition of $z$ together can be treated as
a system of first order autonomous equations
\begin{eqnarray}
  \dot{z} &=& -\frac{m^2\phi}{\sqrt{A}} +
  \frac{\sqrt{3}}{2\mpl}\left( -\sqrt{2} z + 2 \alpha_0 \si (z) \right) \label{zdot}
  \sqrt{z^2+m^2\phi^2-2M} \,, \\
  \dot{\phi} &=& \frac{z}{\sqrt{A}} \,.
\end{eqnarray}
With the equation of state of the field $p_\phi/\rho_\phi$ written in terms of these variables as
\begin{equation}
  \omega_\phi = \frac{2M+z^2-\sqrt{8}\alpha_0|z|-m^2\phi^2}
  {-2M+z^2+m^2\phi^2} \,.
  \label{omegaphi}
\end{equation}

\begin{figure}[htbp]
  \includegraphics[width=80mm]{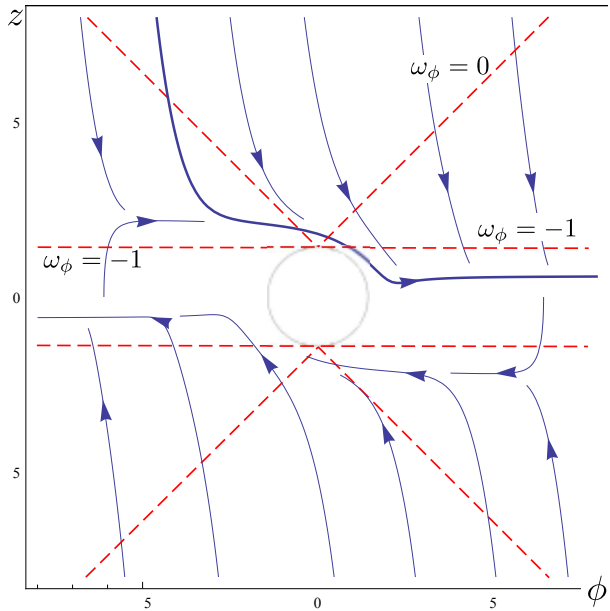}
  \caption{Phase space for the model, the continuous (blue) lines correspond to
  the evolution of the system. The dotted (red) lines correspond to lines of constant
  equation of state, both horizontal lines corresponding to $\omega_\phi=-1$ and
  the four $45^{\circ}$ segments correspond to $\omega_\phi=0$. A typical solution represented
  by the thick line approaches first to $\omega_\phi \sim -1$ (inflation),
  then passes through $\omega_\phi \sim 0$ emulating dark matter and finally comes back
  to $\omega_\phi \sim -1$ at late times as dark energy.}
  \label{fig:phase}
\end{figure}

It can be seen that the system doesn't have any critical points which means that
both variables $z$ and $\phi$ are always time dependent. The usual dynamical system
analysis based on the fixed points is thus not possible in this case. However
the system can be solved numerically to obtain its phase space, shown in Fig. 
\ref{fig:phase}. There we have plotted in dotted (red) lines those of constant equation of
state, the horizontal lines corresponding to $\omega_\phi=-1$ and diagonal lines to $\omega_\phi = 0$.  As can be
seen the sector of initial conditions with big negative $\phi$ values and positive
$z$ values evolves towards a solution with equation of state
near $-1$ which in the phase space corresponds to the left horizontal branch.
This in the unification models is interpreted as the initial period of inflation
in which the equation of state of the solution gets close to $-1$.
To see this one can show that equation (\ref{zdot}) drives $z$ to small values when the constant parameters are in the intervals 
giving in Eq. (\ref{constants}). For the expected value of the field at the
beginning of inflation $\phi_i\sim 15 \mpl$ as obtained in Ref. \cite{DeSantiago:2011qb} under slow-roll conditions, the
value of $z$ will be around $10^{-7}\mpl^2$ corresponding to $\omega\sim -0.994$.
If the system starts in bigger $z$, equation (\ref{zdot}) will acquire negative values
driving the system to this small value corresponding to a potential dominated phase.
For more details on the inflationary realization of the model
see \cite{DeSantiago:2011qb}.

This solution later crosses the lines corresponding to a equation of state equal to 0  
(diagonal lines) that in the unification models correspond to the matter
domination epoch.
The time that the system stays in the regime of $\omega_\phi\sim 0$ has to be long
in order to represent the Dark Matter. This time will depend on the value of the parameters
(\ref{constants}) in the Lagrangian, and in references \cite{DeSantiago:2011qb}
and \cite{Bose:2008ew} it is shown that these parameters can be adjusted in order to obtain this
behaviour from a redshift of order $10^{10}$ up until a recent time when the transition to
$\omega_\phi <0$ has to occur; e.g. Eqs. (\ref{constants}) provide the parameters  for  model $\alpha=1$.
Finally, the solution evolves towards a second period of $\omega_\phi$
close to $-1$, that in the phase space corresponds to the right horizontal branch.
The whole behaviour occurs also for solutions beginning with big positive values of
$\phi$ and negative values of $z$, in which solutions go from positive to
negative values in $\phi$ and live in the $z<0$ part of the phase space, as can be
seen in Fig. \ref{fig:phase}.

The analysis of this phase space is important in understanding the dynamics of
the cosmological solutions of the system (\ref{eq1}, \ref{eq2}). We can conclude 
that an important sector of the possible initial conditions can give
rise to the behavior needed to unify the phenomena of dark matter, dark energy, and
inflation. If fact, half of the possible initial conditions give rise to the
behaviour needed for unification. As was stated in the previous paragraph, all the
solutions that at early times begin with big negative values of $\phi$ and positive values of $z$,
or those that begin with big positive values of $\phi$ and negative values of $z$, achieve 
a successful unified behaviour. A problem that can be seen in Fig. \ref{fig:phase}
is the crossing of the $\omega_\phi=-1$ line, that has been argued in \cite{Vikman:2004dc}
it presents stability problems for k-essence scalar fields, however this will occur
in a future epoch for the case of our Universe, as the current equation of state for the
field is expected to be close to $\omega_\phi=-0.75$
\cite{Bose:2008ew}.

\subsection{Purely kinetic Lagrangian}

The kinetic term in the previous Lagrangian (\ref{eq1})  
was proposed originally in \cite{Chimento:2003ta}
as a purely kinetic Lagrangian, corresponding in our case to the scenario
in which
the potential term (\ref{eq2}) is small compared to the kinetic term
(\ref{eq1}). In this case
the system poses a shift symmetry $\phi \rightarrow \phi+ \phi_0$ which implies that it has
only one degree of freedom. This simplification makes
it possible to obtain an analytical solution to the system \cite{2012arXiv1204.2181S}.
The  energy density and pressure take the form
\begin{eqnarray}
  \rho &=&  (AX)^\alpha - M \,, \label{purerho} \\
   p &=&  \frac{1}{(2\alpha-1)} \left[ (A X)^{\alpha}  
   - 2\alpha  \alpha_0  \sqrt{AX} \right] + M\,. \label{purep}
\end{eqnarray}
Thus, the continuity equation can be written as
\begin{equation}\label{conti}
   \frac{d(AX)}{dN}=\frac{6}{2\alpha-1}(AX)(\alpha_0 (AX)^{1/2-\alpha}-1) \,.
\end{equation}
And defining $y\equiv (AX)^{\alpha-1/2}$, this equation  can be transformed into: 
\begin{equation}
   \frac{dy}{dN} =  3(\alpha_0-y) \, , 
\end{equation}
This differential equation has the analytical solution $y=\alpha_0+c \, e^{-3N}$, where $c$ is a constant of
integration.  After a few e-folds of expansion the constant term dominates.
The equation of state can be computed as
\begin{equation}
   \omega_\phi =\frac{  (A X)^{\alpha}  
   - 2\alpha  \alpha_0  \sqrt{AX}  + (2\alpha-1) M }{ (2\alpha-1)[(AX)^\alpha - M] } \,.
\end{equation}

From the Eq. (\ref{conti}) we can obtain the critical values of the system as
$AX_1=0$ and $AX_2=\alpha_0^{2/(2\alpha-1)}$, which correspond to systems with equation of state
$-1$ in both cases.
To study the stability  we expand equation (\ref{conti}) around each critical point. For the first one, we expand  
$AX=0+\epsilon$ with $\epsilon \ll 1$, then the evolution equation can be approximated as
\begin{equation}
  \frac{d(AX)}{dN} \approx \frac{6\epsilon}{1-2\alpha} \,,
\end{equation}
for $1-2\alpha>0$, which corresponds to a unstable point. If instead $2\alpha-1>0$ then 
\begin{equation}
  \frac{d(AX)}{dN} \approx \frac{6\alpha_0 \epsilon^{(3-2\alpha)/2}}{2\alpha -1 } \,,
\end{equation}
and the critical point is stable for $\alpha_0<0$ and unstable for $\alpha_0>0$. For example,
for the case studied in the equations (\ref{constants}) where $\alpha=1$ and $\alpha_0>0$, the
critical point $X=0$ is unstable. For the cases  in which $X=0$ is unstable, the value of the late time cosmological constant is not 
$\rho(X=0)=M$, that corresponds to the constant added in the Lagrangian.

For the other critical point, we expand as 
$AX=\alpha_0^{2/(2\alpha-1)} + \epsilon$,  then
\begin{equation}
  \frac{d(AX)}{dN} \approx -3\epsilon \,, 
\end{equation}
corresponding to a stable point.
This result is important because the 
dynamical evolution of the system will drive the field to a behaviour
similar to a cosmological constant at late times, with $\omega_{\phi}=-1$ in which 
the system tends to a value of the field with $AX_2=\alpha_0^{2/(2\alpha-1)}$
corresponding to a density $\rho(X=X_2)=\alpha_0^{2\alpha/(2\alpha-1)}-M$ that is
a combination of constants that yields the dark energy of the model, see \cite{DeSantiago:2011qb}. 

\section{Conditions for UDM}

In the first section we studied the model (\ref{eq1}, \ref{eq2}), which
in the late time Universe can give rise to the phenomena of dark matter and
dark energy. However this model is very specific, and therefore the aim of this section is to study  
the conditions for a more general class of scalar fields to reproduce
the same dynamical features. If the Lagrangian has a general form in terms of the field and
the kinetic term $\lag = \lag (X,\phi)$, its equation of state turns out to be
\begin{equation}
  \omega = \frac{\lag}{2X\lag_X-\lag},
\end{equation}
and the sound speed
\begin{equation}
  c_s^2 = \frac{\lag_X}{2X\lag_{XX}+\lag_X}.
\end{equation}
A sufficient condition for the field to behave as dark matter is that both quantities be close to zero
\cite{Bertacca:2010ct}, leaving to the conditions
\begin{equation}
  \frac{\lag}{X\lag_X} \ll 1 \,,
  \label{condeos}
\end{equation}
and
\begin{equation}
  \frac{\lag_X}{X\lag_{XX}} \ll 1 \, .
  \label{condvel}
\end{equation}

There are several Lagrangians that accomplish the above conditions and they have 
been proposed as models for unified dark matter (UDM) meaning that
they can behave as dark matter and, adding a constant to the Lagrangian,
as a combination of dark matter and dark energy. To proceed testing different Lagrangians
we first consider condition (\ref{condeos}), to later analyze models with   (\ref{condvel}).

An example from the
literature is the purely kinetic Lagrangian proposed by Scherrer on Ref. \cite{Scherrer:2004au}
corresponding to
\begin{equation}\label{sche}
\lag = F(X) = F_0 + F_m(X-X_0)^2 \,.
\end{equation}
When the kinetic term is near the minimum $(X -X_0)/X_0 \ll 1$,
this Lagrangian is known to behave as UDM.
 Another similar example of this type is
the ghost condensate model  \cite{ArkaniHamed:2003uy} that served as an attempt to stabilize  k-essence scalar fields
when having an equation of state $\omega_\phi$ smaller than $-1$. These examples show that the behaviour around
the minimum is important.  Let us analyze the conditions for a general Lagrangian having a minimum.
In this case it can be expanded as
\begin{equation}
  \lag (X,\phi)= \lag_0 + \frac{1}{2}\lag_2 \delta^2 + \frac{1}{3!}\lag_3 \delta^3 +
  \cdots \,,
  \label{expansion}
\end{equation}
where $\lag_i$ is the $i$th derivative with respect to $X$
evaluated at the minimum $X_0$, and $\delta$ is the deviation from the minimum, 
$\delta = X- X_0$. The constant term $\lag_0$ can be dropped from the analysis as can be
considered as a cosmological constant. This leave us with the condition
(\ref{condeos}) written as
\begin{equation}
  \frac{\delta}{2X_0} - \frac{6\lag_2+X_0\lag_3}{12X_0^2\lag_2} \delta^2 + \cdots
  \ll 1 \, ,
  \label{condmin}
\end{equation}
that imposes the condition on $\delta/X_0$ to be small, in other words the
deviation from the minimum has to be small.  The 
higher order coefficients in the expansion are close to zero as long as the
first one is, except for very particular cases. In
ref. \cite{Giannakis:2005kr} it is concluded that for Scherrer's model
this deviation $\delta/X_0$ has to be smaller than $10^{-16}$ at the present epoch
to avoid discrepancies in the structure formation and CMB power spectrum in comparison 
with the observations. 

In ref. \cite{DeSantiago:2011qb} it is shown that for the model (\ref{eq1})
the deviation $\delta/X_0$ is of order $10^{-13}$ 
during the equality epoch and at the present
epoch of order $10^{-23}$, resulting in a correct description of the cosmology. 
Considering our  model in more detail, the pressure and density of the purely kinetic part, as given in (\ref{purerho}) and (\ref{purep}), turn  out to be 
around the minimum $AX_0= \alpha^{2/(2\alpha-1)}$ as
\begin{eqnarray}
  P &\approx & M - \alpha_0^{2\alpha / (2\alpha - 1)} +
  \frac{A^2\alpha \alpha_0^{(2\alpha-4)/(2\alpha-1)}}{4} \delta^2 \,, \\
  \rho &\approx &  M - \alpha_0^{2\alpha / (2\alpha - 1)} +
  A\alpha \alpha_0^{(2\alpha-2)/(2\alpha-1)}\delta \, .
\end{eqnarray}
Initially, the terms with $\delta$'s dominate over the constant terms and the effective equation of state of the k-fluid 
behaves as dark matter:
\begin{equation}
  \omega_{\phi} = \frac{A \delta}{4\alpha_0^{2/(2 \alpha -1)}} \approx 0\,.
\end{equation} 
Later on, as $\delta \rightarrow 0$, the solution tends to the attractor in which the constant terms act 
as a cosmological constant, $\rho_\Lambda = -M +\alpha_0^{2\alpha/(2\alpha-1)}$. 

\bigskip 

On the other hand, the condition (\ref{condvel}) states that the speed of sound has to be
small in order to have the growth in the matter inhomogeneities needed for the
structure formation.
For the field around a minimum the condition is written as
\begin{equation}
  \frac{\delta}{X_0} - \frac{2\lag_2+\lag_3}{2X_0^2 \lag_2}\delta^2 + \cdots
  \ll 1 \,. \label{condmin2}
\end{equation}
This condition is similar to (\ref{condmin}) and, except for very particular
Lagrangians, the accomplishment of the first one will be enough.
In other words the deviation from the minimum $\delta/X_0$ must be small.  

A different set of Lagrangians that fulfill (\ref{condeos}) without the necessity of
being around the minimum, are those with a large derivative $d\lag /dX$, for example if
$\lag = AX^\eta$, and $\eta\gg 1$. The condition (\ref{condvel}) for these powerlaw
Lagrangians corresponds to $\eta +1 \gg 1$ that is satisfied once the first condition is.

Another important case is the canonical scalar field $\lag = X - V(\phi)$, where
the condition (\ref{condeos}) becomes $V(\phi)/X \approx 1$. It can be accomplished for
different initial conditions of the field. The 
potential $V(\phi)=\Lambda/2[Cosh^2(\sqrt{3}\phi/2)+1]$ has been studied
in Ref. \cite{Bertacca:2010ct} due to its property of satisfy this condition.
However, the condition (\ref{condvel}) corresponding to have a small speed of sound is
never satisfied as $c_s=1$, making it a bad UDM model. However see \cite{Magana:2012ph}
for arguments in favour of the validity of the canonical scalar field as UDM.

\section{Conclusions}
We have presented a phase space analysis for the unified model (\ref{eq1}, \ref{eq2}) of dark matter, dark energy, and inflation.  We have  
shown that for a large set of the initial conditions ($\phi, \dot{\phi}$) a viable dynamics occurs in which inflation ($\omega_\phi=-1$) happens first, followed by a period of dark matter domination ($\omega_\phi=0$), to finish with dark energy ($\omega_\phi=-1$).  An intermediate 
radiation period is possible in this model once it is added an extra radiation component as in the 
standard model of cosmology.

Once inflation ends, the model is fully described by the purely kinetic Lagrangian
$\lag=F\left( X \right)$ with $F$ as in Eq. (\ref{eq1}). We have shown 
that this system possesses  a late time stable solution in which  $\omega_{\phi} = -1$, that is dark energy.   In ref. 
\cite{DeSantiago:2011qb} the range of  parameters were given to achieve a successful cosmological model, and in the present work 
the dynamical analysis clearly shows why the system is tenable.
A problem however may arise in the form of possible instabilities due to the crossing of
the system trough $\omega_\phi=-1$, as observed in Fig. \ref{fig:phase}, but this crossing will occur
in the future as the equation of state should have only
moved from $\omega_\phi \sim -1$ in the matter dominated epoch to $\omega_\phi\sim -0.75$
in the present epoch.

In the last part of our work we have presented the general features that are necessary to have a model that behaves as dark matter. If one adds 
a cosmological constant to this model, one ends with a unified dark matter and dark energy model, called generically UDM.  There are two
conditions that these models should fulfill, equations (\ref{condeos}, \ref{condvel}) playing the role of an effective fluid with small 
pressure (in comparison  to the density) and small speed of sound (in comparison to the speed of light).     We have analyzed some 
models studied in the literature that fulfill these conditions. In particular, $F(X)$ models that possess a
minimum, as Sherrer's model (\ref{sche}) or
the model studied in the first section (\ref{eq1}),  when they are close enough to the minimum,
they behave as dark matter.    Departures from the minimum 
cause a change in the transfer function and therefore to a different
growth history in comparison to the standard model of 
cosmology \cite{Giannakis:2005kr}.  Additionally, we have analyzed other models of the literature, such as models
with large  derivative $\lag_X$, for example $\lag=AX^\eta$ with large $\eta$, and models with canonical Lagrangian
$\lag =X-V(\phi)$. The constrictions were given for these models.

\begin{theacknowledgments}
We gratefully acknowledge support from  CONACYT Grant Nos. 210405 and 84133-F. We thank the referee for valuable comments.  
\end{theacknowledgments}

\bibliographystyle{aipproc}   

\bibliography{biblio}{}

\end{document}